# Auroral Morphological Changes to the Formation of Auroral Spiral during the Late Substorm Recovery Phase: Polar UVI and Ground All-Sky Camera Observations

**Motoharu Nowada[1][†], Yukinaga Miyashita[2,3], Noora Partamies[4,5],**

**Alexander William Degeling[1], and Quan-Qi Shi[1]**

[1] Shandong Key Laboratory of Optical Astronomy and Solar-Terrestrial Environment, School of Space Science and Physics, Institute of Space Sciences, Shandong University, Weihai, Shandong, People's Republic of China.

[2] Korea Astronomy and Space Science Institute, Daejeon, South Korea.

[3] Korea University of Science and Technology, Daejeon, South Korea.

[4] Department of Arctic Geophysics, The University Centre in Svalbard, Norway.

[5] Birkeland Centre for Space Science, Norway.

[†]Corresponding author: Motoharu Nowada (moto.nowada@sdu.edu.cn)

**Key Points:**

- Auroral spiral was formed while the substorm-associated auroral bulge was subsiding and poleward-grown auroral streak structures appeared.

- The source region of the auroral spiral in the magnetotail is widely distributed over 30 $R_E$, from $X_{gsm} \sim$ -40 to -70 $R_E$.

- Extensive areas of the magnetotail are sufficiently active to cause auroral spirals even during the late substorm recovery phase.

**Running Title:** Auroral Spiral in Late Substorm Recovery

## Abstract

The ultraviolet imager (UVI) of the Polar spacecraft and an all-sky camera at Longyearbyen contemporaneously detected an auroral vortex structure (so-called "auroral spiral") on 10 January 1997. From space, the auroral spiral was observed as a "small spot" (one of an azimuthally-aligned chain of similar spots) in the poleward region of the main auroral oval from 18 h to 24 h magnetic local time. These auroral spots were formed while the substorm-associated auroral bulge was subsiding and several poleward-elongated auroral streak-like structures appeared during the late substorm recovery phase. During the spiral interval, the geomagnetically north-south and east-west components of the geomagnetic field, which were observed at several ground magnetic stations around Svalbard island, showed significant negative and positive bays caused by the field-aligned currents related with the aurora spiral appearance. The negative bays were reflected in the variations of local geomagnetic activity index ($SML$) which was provided from the SuperMAG magnetometer network at high latitudes. To pursue the spiral source region in the magnetotail, we trace each UVI image along field lines to the magnetic equatorial plane of the nightside magnetosphere using an empirical magnetic field model. Interestingly, the magnetotail region corresponding to the auroral spiral covered a broad region from $X_{gsm} \sim$ -40 to -70 $R_E$ at $Y_{gsm} \sim 8$ to 12 $R_E$. The appearance of this auroral spiral suggests that extensive areas of the magnetotail (but local regions in the ionosphere) remain active even when the substorm almost ceases, and geomagnetic conditions are almost stable.

## Plain Language Summary

Auroras that locally exhibit a vortex structure are referred to as auroral spirals. The fundamental features of auroral spirals, such as their generation process and source region in the magnetotail, are poorly understood. Based on the images obtained from the Polar UVI instrument and an all-sky camera at Longyearbyen, we examined the morphological changes before and during the lifetime of an auroral spiral, which occurred during a comparatively quiet-time of magnetotail recovery following a main substorm phase on 10 January 1997. The auroral spiral was formed after substorm-associated auroral features subsided and the disappearance of several poleward-elongated auroral streak-like structures. During the spiral, the geomagnetic fields observed around Svalbard island and the geomagnetic activity index derived from the high-latitude magnetometers showed significant negative/positive variations related with the aurora spiral appearance.

According to the field-aligned auroral spiral UVI image projections onto the nightside magnetic equatorial plane using an empirical geomagnetic field model, the source region of the auroral spiral spanned a broad region over 30 $R_E$ in the magnetotail direction with a dawn-dusk width of $\sim 4$ $R_E$. These results suggest that extensive areas of the magnetotail are active enough to cause auroral spirals even during the late substorm recovery phase.

## 1. Introduction

Auroral phenomena with vortex structures (auroral spirals) have been frequently observed by ground-based all-sky cameras and polar-orbiting satellites. Their fundamental characteristics, such as their appearance locations and spiral diameters, have been discussed by Davis and Hallinan (1976), Partamies et al. (2001a, b), Voronkov et al. (2000), and references therein. Partamies et al. (2001a) and Davis and Hallinan (1976) independently obtained statistical distributions of auroral spirals in magnetic local time (MLT) and magnetic latitudes (MLat), based on all-sky camera observations. The previous studies, however, neither discussed nor examined how auroral spirals are formed, for example, by following successive space- and ground-based auroral images. Besides spirals, two other types of auroral arc accompanied by vortex structures, known as curl and fold, were identified by Davis and Hallinan (1976) and Hallinan and Davis (1970). According to these studies, spirals have the largest scale (> 50 km). Curls are small-scale (< 10 km) vortices embedded in auroral arcs, and folds are distorted arcs with an intermediate scale size ($\sim 20$ km). The rotational senses of spirals and folds are counter-clockwise when viewed from above in the Northern Hemisphere. In contrast, curls are clockwise. The horizontal velocities of folds and curls range from 0 to 5 km/s and from 0 to 22 km/s, respectively. All three types of vortical arc structures can appear simultaneously in the same auroral form and may not be easily distinguishable. Folds often change into curls, and curls and spirals have the same general appearance.

Most auroral spirals were found to be distributed over the nightside ionosphere from 18 h to 5 h MLT, while their MLat distribution was concentrated around 65° (Partamies et al., 2001a), or between 70° and 80° (Davis and Hallinan, 1976), although there were constraints on fields of view due to the all-sky camera locations in these studies. Depending on the study, the scale (diameter) of the auroral spiral varies between 25 and 75 km (Partamies et al., 2001a, 2001b), or between 20 and 1,300 km (Davis and Hallinan, 1976).

Keiling et al. (2009a, 2009b) examined an auroral spiral event that occurred during an auroral substorm, from the onset to the expansion phase; this event was associated with fast earthward plasma flows, an abrupt increase of the northward magnetic field (dipolarization), and energetic particle injections, based on in situ magnetotail observations. These authors obtained a series of images of auroral development from auroral breakup (onset of poleward expansion of the aurora) to auroral spiral decay using the THEMIS all-sky images. Because the spirals were observed during the substorm expansion phase, they were formed and decayed in conjunction with repetitive shrinking-and-stretching of multiple arcs associated with substorm growth, and in the westward travelling surges (e.g., Amm and Fujii, 2008; Marklund et al., 1998; Roux et al., 1991). The complete auroral spiral was short-lived (less than 10 s), and the time scale from formation to decay was ~ 1 min. The average diameter of the observed spiral was 200 – 300 km.

However, as pointed out by Partamies et al. (2001a), the occurrence probability of auroral spirals is higher in geomagnetically quiet times without substorm activity, including during the substorm recovery phase, than during the substorm onset and expansion phases. In their statistics, geomagnetically quiet time was defined by the local magnetic north-south component of geomagnetic field less than 200 nT. However, it remains unclear how shape of auroral spirals (arcs) can be formed under how geomagnetic conditions, such as during the non-substorm interval, or early and late substorm recovery phases. The auroral morphological changes to the complete auroral spiral formation and the source region of the auroral spiral in the magnetosphere have not yet been examined and reported. Keiling et al. (2009a, 2009b) concluded that auroral spirals are produced by field-aligned currents (FACs) initiated by flow shears (vortices) between magnetotail reconnection-generated fast plasma flows and slower background plasma sheet flows. However, auroral spirals exhibit not only substorm activity dependence but also relatively broad MLT and MLat distributions (Davis and Hallinan, 1976; Partamies et al., 2001a). Therefore, the source region in the magnetotail and/or formation mechanism of all auroral spirals are most likely not always the same. Lysak and Song (1996) focused on that magnetosphere-ionosphere coupling played a significant role in auroral spiral formation, and stated that the current sheet instability (CSI) in the nightside magnetosphere is a key mechanism for the spiral formation. CSI has the same physical characteristics as the Kelvin-Helmholtz instability (KHI) but is formed by magnetic shear generated by upward (from the ionosphere to the magnetosphere) FACs due to the velocity shear. The rotational sense of auroral spirals (counter-clockwise when viewed from above in the Northern Hemisphere) is opposite to the Kelvin-Helmholtz vortex (clockwise or curls). The auroral spiral can finally be formed by CSI, caused by magnetic shear under the electric conductance difference between the ionosphere and the magnetosphere.

In this paper, we examine the morphological changes of auroral spirals which occurred prior to and during the late substorm recovery phase observed on 10 January 1997, and the associated variations of the auroral source region in the magnetotail, based on the Polar UVI observations. The instrumentation is described in Section 2. In Section 3, we show the solar wind conditions, and the results of Polar UVI and ground-based geomagnetic field observations. The summary and discussion of this study are described in Section 4. In Section 5, we described the conclusion of this study.

## 2. Instrumentation

The ultraviolet imager (UVI) instrument onboard the Polar spacecraft, launched on 24 February 1996, provides global auroral imaging data (Torr et al., 1995). The UVI images used in this study consist of long Lyman-Birge-Hopfield emission (LBHL; ~170 nm) and short LBH emission (LBHS; ~150 nm) images. The integration times of the Polar UVI data are 18 s and 36 s. In this study, the UVI images are displayed in altitude adjusted corrected geomagnetic (AACGM; Baker and Wing, 1989) and geographic coordinates. It has been well known that the UVI image data are degraded in the direction perpendicular to the Polar spacecraft track by satellite wobble motion (e.g, Parks et al., 1997). However, this wobble of the Polar spacecraft is predictable (Parks et al., 1997). In this study, we used the UVI image data after most of the wobble effects were removed.

The all-sky camera (ASC) installed at the Longyearbyen station (75.32° magnetic latitude and 111.0° magnetic longitude, LYR) covers a circular area with a diameter of about 600 km at 110 km altitude with a field of view of 140°. Because the number of pixels corresponding to a 140° field of view is 440×440, the average spatial resolution at LYR is 1.36 km/pixel (600 km/440 pixels). The time resolution of the ASC's images filtered at 558 nm is 20 s for the time interval of interest. By examining the ASC images on the ground, we can clearly identify whether or not auroral spiral and surrounding auroral signatures observed by Polar are discrete or diffuse auroras, and also determine auroral arc types (spirals, curls, or folds).

## 3. Observations

Figure 1 shows the OMNI solar wind parameters and geomagnetic indices for 5 h between 17:00 UT and 22:00 UT on 10 January 1997. From top to bottom, the panels show the *AL* and *AU* indices, the GSM-Y and -Z components of the interplanetary magnetic field (IMF-$B_y$ and -$B_z$), the IMF clock angle (derived by arctan(IMF-$B_y$/IMF-$B_z$)), and the Akasofu-Perreault parameter ($\varepsilon_{AP}$), which is a measure of the solar wind energy input rate (Perreault and Akasofu, 1978). The *Kp* index is also shown at the bottom of the figure. During this interval, the entire cycle of a substorm driving moderate geomagnetic disturbances with a *Kp* range from 3- to 4 can be found. *AL* shows a sharp, large decrease from ~18:45 UT (the substorm onset) to ~19:05 UT, and then *AL* recovered to ~ 0 nT at ~21:23 UT. The interval of the auroral spiral was identified by visual inspection, based on the data from Polar UVI and the all-sky camera (ASC) installed at the Longyearbyen station, and persisted from 19:59 UT to 21:23 UT (bracketed by two black broken lines with horizontal thick bars) during the late recovery phase of the substorm. Clear jumps were seen in the associated IMF-$B_y$ and -$B_z$ components at 21:10 UT. In particular, IMF-$B_z$ turned from long-lived (at least 4 h) weak southward (negative) to northward (positive) directions, while IMF-$B_y$ had a dominantly dawnward (negative) component. The clock angle became increased from -75° to -25° in association with this abrupt IMF-$B_z$ jump. Accordingly, the $\varepsilon$ parameter shows a significant decrease associated with the jumps of the IMF and clock angle, which was preceded by a gradual decrease during the first half of the spiral interval. Note that the increasing trend of the $\varepsilon$ parameter before the spiral appearance from ~18:10 UT to ~19:15 UT may have caused the substorm.

Corresponding auroral activity and its morphological changes under these solar wind conditions are shown in Figure 2. A series of representative auroral images before the onset of the spiral is shown in Figures 2a and 2b, which correspond to the early and late substorm recovery phases, respectively. The substorm-associated auroral bulge azimuthally extended from 23 h to 4 h MLT along the main auroral oval (Figure 2a), and reduced in size and intensity with time. Several poleward-elongated auroral streak-like structures were found to remain in place as the auroral bulge subsided (Figure 2b). As seen in Figure 2c, the auroras changed their forms from several elongated streak-like structures to azimuthally-aligned four small spots, which are seen from 20 h to 1 h MLT. The spot at ~ 23 h MLT, highlighted with a yellow oval, corresponds to the aurora with significant vortex structure, that is, the auroral spiral seen at Longyearbyen. Such azimuthally-aligned auroral spots in the poleward region of the main auroral oval from the premidnight to midnight sectors have been detected previously with the Viking UV imager, but were not discussed in detail, because they were identified as the poleward component of a "double auroral oval" (Elphinstone et al., 1995). The simultaneous all-sky camera (ASC) images at Longyearbyen are shown in Figures 2g and 2h, in which the auroral spiral is clearly visible in the southern part of the ASC's FOV. The rotational sense of the observed spiral is anti-clockwise, when viewed from the direction of the magnetic field. From the ASC's images in geographical coordinates, we roughly estimated the scales (diameters) of the auroral spirals with various directional patterns, and summarized their average scales in Table 1. Here, we estimated the two kinds of spiral scale for each time using 1° in geographical latitude = 111 km: the core part (central part of intense aurora) and the core part with the spiral arms. The actual calculations were performed using Latitude/Longitude Distance Calculator, which was developed by National Hurricane Center and Central Pacific Hurricane Center (https://www.nhc.noaa.gov/gccalc.shtml), by giving the geographical latitudes and longitudes at two different edge points of the spiral.

As the substorm recovery proceeded, several intense spots were clearly seen in the poleward region of the main auroral oval (Figures 2d and 2e). Figures 2i and 2j show the ASC's images from the Longyearbyen station for the times nearest to the Polar UVI observation times (20:56:20 UT and 21:02:20 UT). The FOV of the ASC is highlighted with a yellow oval in Figures 2d and 2e. Auroral spirals clearer than the previous spirals (Figures 2g and 2h) can be found. They moved southwestward with anti-clockwise rotation, compared with the previous spirals (more clearly seen in Movie S1). The average scale (diameter) of the core part (core + arm part) of the detected spiral was about 150 – 180 km (260 – 320 km). To check whether or not the auroral spot detected by the Polar UVI corresponds to the spiral observed by the Longyearbyen ASC, we compared the plots of the UVI data in geographical coordinates (Figure 2f) with the spiral images observed at Longyearbyen (Figures 2g, 2h, 2i and 2j). The auroral spiral (yellow oval) detected by Polar covers the southern region of Longyearbyen on Svalbard island, demonstrating that the auroral spiral seen by the Polar UVI is identical with that detected by the ASC at Longyearbyen. We show the UVI data in two different coordinates: altitude adjusted corrected geomagnetic (AACGM) (Figure S1) and geographic coordinates (Figure S2) from 18:41:28 UT to 21:57:44 UT, which include the whole auroral spiral interval.

Figure 3 shows four zoomed-in LBHL-36s UVI images of the auroral spiral in the nightside ionosphere from 18 h to 6 h MLT. The color code is assigned according to the logarithm of auroral brightness in units of Rayleigh. The time sequence of these images is the same as that of Figure 2f. The blue circle indicates the auroral spiral which was detected by the ASC at Longyearbyen. The four bright spots were azimuthally aligned in the poleward part of the main auroral oval from 20 h to 1 h MLT (Figures 3a and 3b). However, from these UVI images alone, we cannot identify whether or not these spots are aligned spirals simulated and observed by Huang et al. (2022) or four independent auroral spirals. Taking a look at Figures 3c and 3d, the number of the spots decreased to one or two, but the auroral spiral seen at Longyearbyen remained. From a series of the images, the auroral spiral detected by the ASC at Longyearbyen moved eastward slowly, and approached the midnight sector, changing its shape. We cannot clearly identify the spiral shape from the UVI images, because the Polar orbital altitude was too high for the UVI instrument to obtain image data with sufficiently high spatial resolution.

The series of the spiral's dynamical morphological changes and its eastward motions to the midnight sector were also detected by the ASC at Longyearbyen. Figure 4 shows the zoomed-in Polar UVI images of Svalbard (Longyearbyen) and its surrounding regions as shown in Figure 2f (Figures 4a – 4d) and the corresponding snapshots of the Longyearbyen ASC data plots in geographic coordinates (Figures 4e – 4h). In the ASC plots, we assumed that the auroral emission height was 110 km. Vertical and horizontal axes give the geographical latitude and longitude, respectively, in units of degree (°). The color code is assigned according to the auroral luminosity in units of Rayleigh. The red circle in each panel indicates the location of the Longyearbyen station. We cannot discuss the detailed structure of the spiral, based on the comparison between the Polar UVI and ASC data. The whole spiral shape, however, seems to correspond well to each other. The spiral has a bright circular shape in panels a and e. The elliptical but less brighter spiral was seen in panels b and f. In panels c and g, and d and h, the spiral brightness became more intense, and its shape was stretched more equatorward. This comparison also guarantees a good correspondence between the Polar UVI and the ASC imager observations by showing the geographical latitudes and longitudes of the auroral spiral at Longyearbyen.

Figure 5 shows the geomagnetic activity indices (the *SML* and *SMU* indices) and geomagnetic field data obtained from the five magnetometers installed at and around Longyearbyen station (LYR) on Svalbard island (the International Monitor for Auroral Geomagnetic Effects (IMAGE) magnetometer array, also see the details in Tanskanen, 2009) during 5 h from 17:00 UT to 22:00 UT. From top to bottom, the panels display the *SML* and *SMU* indices, and the three components of the magnetic field ($B_N$, $B_E$, and $B_Z$) at each of the five magnetometer stations. The N, E, and Z directions correspond to geomagnetic north-south, east-west, and up-down, respectively. These geomagnetic observatories have been sorted in decreasing order of latitude. The auroral spiral interval from 19:59 UT to 21:23 UT is indicated with a horizontal gray arrow and also bracketed with gray broken vertical lines.

When the spiral was observed, the substorm was nearly recovered and geomagnetic conditions were also nearly stable. The geomagnetic fields around Longyearbyen (where the spiral appeared), however, showed significant perturbations in all magnetic field components. In particular, clear negative excursions (bay variations) were seen in the $B_N$ components at all stations and the $B_Z$ component at BJN, and clear positive bay variations of $B_Z$ were also observed at NAL, LYR, and HOR (HRN) stations at ~20:15 UT and ~21:12 UT. Although these negative bay variations are reflected in the variations of the *SML* index, the usual *AL* as shown in Figure 1 showed no significant negative variations. This suggests that the field-aligned currents (FACs) associated with the auroral spiral were concentrated around Longyearbyen in this case.

Although the auroral morphological changes to the spiral are clarified, we do not have any observations of the source region of the spiral and its temporal and spatial variations. To pursue the source region of the auroral spiral and its variations, we made magnetic equatorial plane projection maps, where each pixel of the Polar UVI images was traced to the GSM X-Y plane of the magnetotail along the magnetic field lines, based on the Tsyganenko 96 model empirical magnetic field model (T96, Tsyganenko and Stern, 1996). Actual UVI image mapping to the magnetic equatorial plane using the T96 model was performed, based on the calculation routine implemented in the Space Physics Environment Data Analysis Software (SPEDAS) 4.1 package (Angelopoulos et al., 2019). In this study, however, we set up the mapping algorithm to remove the limitation of the maximum trace distance that was set in the model: the radius of a sphere, defining the outer boundary of the tracing. Figure 6 shows the projection maps of the UVI images onto the GSM X-Y plain for times corresponding to the auroral bulge (Figure 6a), poleward-elongated streak-like structures (Figure 6b), and auroral spirals (Figures 6c to 6f). These times are the same as those of Figure 2f. Such mappings to the magnetic equator have already been established and implemented by many researchers (e.g., Elphinstone et al., 1993, Lu et al., 1999, 2000, Pulkkinen et al., 1995, 1998, and references therein). Note that these mapping results of the auroras to the equatorial X-Y locations may sensitively depend on magnetic field models used, as suggested by Lu et al. (2000; see their Plate 3).

In Figures 6a and 6b, the projection maps before the spiral formation are shown. The azimuthally-extended auroral bulge in the ionosphere is broadly projected onto the magnetotail (Figure 6a). The particularly intensified part (dark red part) extends from $X_{gsm} \sim -10$ to $-90$ $R_E$ and $Y_{gsm} \sim -10$ to $10$ $R_E$. The elongated auroral streak-like structures are also distributed from $X_{gsm} \sim -20$ to $-40$ $R_E$ or, at most, $-70$ $R_E$ in the duskside (Figure 6b). The auroral spiral detected by the ASC at Longyearbyen (Figures 6c to 6f) broadly covers the magnetotail region from $X_{gsm} \sim -40$ to $-70$ $R_E$ at $Y_{gsm} \sim 8$ to $12$ $R_E$, as highlighted by the yellow oval. The projected elongated auroral streak-like structures are "elliptically" distributed from the duskside to the dawnside of the magnetotail. Kaufmann et al. (1990) concluded that all nightside auroral phenomena and nightside FAC structures in the ionosphere which were projected onto the magnetotail, based on the T89 magnetic field models for geomagnetically quiet times (Tsyganenko, 1989), essentially have magnetic equatorial sources and are connected to the plasma sheet or the plasma sheet boundary layer. Furthermore, they revealed that the (circular) ionospheric structure, such as nightside auroral arcs and FACs, are mapped to a highly elongated ellipse over a broad range of X direction in the dawn-dusk magnetotail under geomagnetic quiet conditions. Lu et al. (2000) showed that even during a substorm interval, the ionospheric auroras, which were mapped using the T96 model (Tsyganenko and Stern, 1996), also correspond to highly elongated elliptical structures along the magnetotail on the magnetic equatorial plane. Their results would be supportive of justifying our mapping results.

Note that these UVI data mappings to the nightside equatorial plane cannot be used to specify and discuss either the precise location where some triggering plasma process occurred to form the auroral spiral, or the detailed formation mechanism of the spiral. The mapping can be effective in knowing the magnetotail source region of the spiral, that is, how much extent the spiral was distributed in the nightside magnetosphere.

## 4. Discussion

We examined the auroral morphological changes prior to and during auroral spiral appearances in the late substorm recovery phase, observed on 10 January 1997, and found that the source region of the spirals extends over $30$ $R_E$ from $-40$ to $-70$ $R_E$ downtail with a $Y_{GSM}$ width of $\sim 4$ $R_E$ according to mappings along field lines to the magnetotail. These highly elliptically-elongated structures along the magnetotail on the magnetic equatorial plane are consistent with the results of Kaufmann et al. (1990) under geomagnetic quiet conditions and those of Lu et al. (2000) for a substorm. Recently, using particle-in-cell simulations and low-altitude satellite particle measurements, Huang et al. (2022) showed that an auroral spiral can be formed by magnetic reconnection in the ionosphere. Their spirals appeared as a chain of multiple and azimuthally-aligned structures. Our observational results are not sufficient to discuss how magnetotail process(es) contributed to the auroral spiral formation. Unfortunately, because of low spatial resolution of Polar UVI images from a high altitude and the spiral detection at only a single point (LYR), we also cannot clarify whether or not our spiral really is in fact a member of such a chain of structures. Therefore, it remains unclear what significantly differs between the auroral spiral of Huang et al. (2022) generated by magnetic reconnection in the ionosphere and our spiral that would have the source on the magnetic equatorial plane of the magnetotail.

The substorm-associated bulge, which was azimuthally extended, became smaller as the substorm recovery proceeded, and the auroral morphology changed to several poleward-elongated streak-like structures. After these auroral streak-like structures were formed, the Polar spacecraft detected that several small auroral spots were consecutively and azimuthally distributed in the poleward region of the main auroral oval near midnight. Davis and Hallinan (1976) and Elphinstone et al. (1995) reported an observational case of azimuthally-aligned small auroral spots at the poleward edge of the main auroral oval from premidnight to midnight. Also in this study, Polar detected two or four small azimuthally-aligned spots at the poleward edge of the nightside auroral oval (see Figures 2c and 2e). At least one of them was an auroral spiral as detected by the Longyearbyen ASC, but we cannot conclude that the other small spots were auroral spirals because clear observational evidence to advocate this could not be obtained, based on the Polar UVI images and ASC observations. The other ASCs which are installed at the points around LYR and Svalbard island, such as the Greenland east coastline, the Norwegian west coastline, and the north of Iceland and Russia, clearly detected neither the spiral seen at LYR nor the (chain of) other small spots above the (Barents) sea within their fields of view. Because some ASC locations were also covered by thick clouds, we could not visually identify auroral spirals.

The auroral spots observed by Elphinstone et al. (1995) occurred during the very early stage of the substorm recovery; the associated *AL* still kept its value of ~ -340 nT, and the intense substorm-associated auroral bulge structure still remained in the main auroral oval. These observations suggest that the spots of Elphinstone et al. (1995), which were potentially spirals, may be different from our case in the background environment of the auroral spiral formation.

The auroral spiral observed and discussed in this study and the auroral spots found by Elphinstone et al. (1995) were auroral phenomena which brightened at the poleward edge of the nightside main auroral oval. Therefore, these might be poleward boundary intensifications (PBIs) (e.g., Lyons et al., 1998, Zesta et al., 2002, and references therein). The PBIs sometimes evolve into vortical structures and can resultantly be seen as "spiral" (e.g., Zesta et al., 2002). The spiral also glows at the poleward edge of the main auroral oval as well as PBIs. However, because, unfortunately, there were no in situ observations in the magnetotail in this spiral event, we cannot elucidate whether or not our spiral was caused by the same generation mechanism as PBIs, such as distant magnetic reconnection and associated fast earthward plasma flows in the plasma sheet (e.g., de la Beaujardière et al.,1994, Lyons et al., 1999, Zesta et al., 2000, 2006, and references therein). Using the T96 model, Zesta et al. (2000) roughly estimated that the scale of PBIs reached -50 – -100 $R_E$ in the X direction in the magnetotail, nearly consistent with the X-directional scale of our spiral.

We, here, briefly discuss the difference between the present spiral, and other auroral phenomena, such as auroral streamers, auroral beads, and the omega band aurora. Streamers, which approximately orient to the north – south direction, appear after PBIs and frequently propagate equatorward (e.g., Ebihara and Tanaka, 2016, Nishimura et al., 2011, 2013, and references therein). As Kepko et al. (2009) and Nishimura et al. (2010) pointed out, the auroral streamers (referred to as N-S arcs) may sometimes lead to the substorm onset. They, however, do not evolve to the spiral-type auroras, as shown by Davis and Hallinan (1976) and Hallinan and Davis (1970). Furthermore, the auroral streamers often appear prior to the substorm onset, but our spiral was seen during the late stage of the substorm recovery phase. Therefore, the substorm phase is also different between streamers and the auroral spiral.

Auroral beads are also well known as an auroral phenomenon of substorm onset and are frequently observed near the equatorward boundary of the main auroral oval. The auroral beads evolve as periodic waves (undulations) along the arc and enter more dynamic and nonlinear evolution stage, leading to the auroral breakup. According to ground- and satellite-based measurements performed by Motoba et al. (2012) and Gallardo-Lacourt et al. (2014), the auroral beads are caused by the plasma instabilities in the nightside plasma sheet at onset. These instabilities would not be a pure MHD scale process but include kinetic effect (Motoba et al., 2012;

Gallardo-Lacourt et al., 2014). On the other hand, the spiral observed clearly had a longitudinal spot structure at the poleward edge of the main auroral oval during the late substorm recovery phase, but we cannot identify whether or not the auroral spiral was associated with an arc, only based on the Polar UVI observation. The spiral morphology kept spot-like shape but was not annihilated by such as nonlinear evolution. Motoba et al. (2012) tried to project the auroral beads detected by the all-sky images onto the magnetotail equatorial plane using the T96 model as we performed in this study. Their auroral beads were traced in a narrow magnetotail region on the pre-midnight near-Earth tail equatorial plane X ~ -7.75 to -8.25 $R_E$ and Y = 1.0 to 2.0 $R_E$, supporting that there is significant difference in the appearance locations between the auroral beads and the present spiral. The scale length of each auroral bead segment in the Y direction was 500 – 800 km and seems to be larger than the spiral in this case, but the X-Y ranges of the projections onto the magnetic equatorial plane in the magnetotail are much smaller than spiral discussed in our study, suggesting that the auroral spiral is different from auroral beads.

The omega band aurora can appear around lower MLat from 65° to 70°, which correspond to the equatorward region of the auroral oval (e.g., Apatenkov et al., 2020; Liu et al., 2018). The auroral spiral, however, can be seen as spot at the poleward edge of the main auroral oval, supporting that the observed spiral is different from the omega band aurora. The omega band aurora appears and evolves in association with the wave(-like) structures, but the spiral neither has wavy structure nor shows significant development in wave.

Keiling et al. (2009a) also observed an auroral spiral, but their case was associated with auroral arc development and a westward traveling surge during the substorm expansion phase, unlike our auroral spiral during the late substorm recovery phase. Furthermore, the duration of our auroral spiral is different from that of Keiling's spiral case. Murphree and Elphinstone (1988) and Keiling et al. (2009a) pointed out that short-lived (less than 1 min) auroral spirals are expected to occur frequently during the (early and late) substorm recovery phases, and auroral spirals are one of the well-known representative short-lived auroral phenomena. However, although the ASC at Longyearbyen detected repetitive formations and decays of auroral spirals, a single spiral with a duration much longer than the previous cases was also clearly seen (see Movie S1). Therefore, this observed spiral might differ from the short-lived auroral spiral during the substorm expansion phase that Keiling et al. (2009a) reported.

The specific relationship between spiral occurrence and IMF orientation/variation still remains an open question. Although the IMF-B$_z$ component changed its polarity from weakly negative to positive in the latter half of the spiral interval, no significant changes can be found in the spiral profile. The IMF clock angle was almost constant ($\theta_{CLOCK} \sim -95°$) prior to and during the spiral, suggesting that our auroral spiral might have occurred under weak but persistent large-scale plasma convection. Any relationship between magnetic reconnection (rate) maintaining large-scale plasma convection and fundamental formation process of the spiral remains unclear. Elphinstone et al. (1995) and Murphree and Elphinstone (1998) discussed auroral spiral cases under northward IMF-B$_z$ or IMF-B$_z \sim 0$ nT. Considering only the IMF conditions, our auroral spiral should be different from their cases. Partamies et al. (2001b) suggested that the winding of the auroral spiral can be attributed to local plasma processes, such as FAC enhancements. There are several essential problems to be elucidated, in particular: i) by what plasma phenomena in the magnetotail are auroral spirals basically formed; ii) why do they have locational and substorm phase dependences; and iii) how do spiral formation processes differ, depending on locations and substorm phases.

In this study, we cannot discuss the detail auroral spiral formation mechanism, based on our results alone. Furthermore, it is hard to compare our results with the formation processes addressed by theories and computer simulations, such as CSI in the nightside magnetosphere (Lysak and Song, 1996) and protuberance of a high-pressure region into up- an downward FAC pair from the nightside plasma sheet (Ebihara and Tanaka, 2016), as discussed within a framework of magnetosphere-ionosphere coupling. This is attributed to absence of in situ magnetotail observational data corresponding to the spiral occurrence. More detailed discussion and comparison between in situ spiral observations and a series of the spiral theories/simulations are future works.

## 5. Conclusions

In this study, we first detected the auroral spots (spirals) that were formed during the late substorm recovery phase. Projecting the auroral spiral along field lines onto the magnetic equatorial plane of the magnetotail using an empirical geomagnetic field model, the spiral source region was found to be extensively distributed from -40 to -70 R$_E$ downtail. This result suggests that extensive areas of the magnetotail are active enough to cause auroral spirals even during the late substorm recovery phase.

## Acknowledgments

M.N. enjoyed fruitful and constructive discussions with Jong-Sun Park and Timo Pitkänen, and was supported by a grant of the National Natural Science Foundation of China (NSFC 42074194). Y.M. was supported by Korea Astronomy and Space Science Institute under the R&D program (2022-1-850-09) supervised by the Ministry of Science and ICT. N.P. was supported by the Norwegian Research Council (NRC) under CoE contract 223252. Q.Q.S. was supported by NSFC 41731068, 41961130382, and 41974189, and also supported from International Space Science Institute, Beijing (ISSI-BJ). We thank George K. Parks for providing the Polar UVI data and Kan Liou for processing the data.

## Data Accessibility

Polar ultraviolet imager (UVI) level-1 data can be accessed from https://cdaweb.gsfc.nasa.gov/pub/data/polar/uvi/uvi_level1/. Data for calibrating the level-1 data and calculating the position of the UVI images can be accessed from https://doi.org/10.6084/m9.figshare.5197084.v1 (Uritsky and POLAR UVI team, 2017). Magnetometers – Ionospheric Radars – All-sky Cameras Large Experiment (MIRACLE) ASC quick-look data are available at https://space.fmi.fi/MIRACLE/ASC/, and ASC full resolution images and their numerical data used in this study can be downloaded from https://doi.org/10.5281/zenodo.6552492 (Nowada et al., 2022). All numerical data of the magnetometers in the International Monitor for Auroral Geomagnetic Effects (IMAGE) geomagnetic observatory chain, which are used in this study, can be downloaded from https://space.fmi.fi/image/www/index.php?page=user_defined, and the SuperMAG network website (https://supermag.jhuapl.edu/mag/). We thank the institutes that maintain the IMAGE Magnetometer Array: Tromsø Geophysical Observatory of UiT the Arctic University of Norway (Norway), Finnish Meteorological Institute (Finland), Institute of Geophysics Polish Academy of Sciences (Poland), GFZ German Research Centre for Geosciences (Germany), Geological Survey of Sweden (Sweden), Swedish Institute of Space Physics (Sweden), Sodankylä Geophysical Observatory of the University of Oulu (Finland), Polar Geophysical Institute (Russia), and DTU Technical University of Denmark (Denmark). The *SML* and *SMU* indices can be accessed from the SuperMAG network website (https://supermag.jhuapl.edu/mag/). The OMNI solar wind magnetic field and plasma data can be acquired from Coordinated Data Analysis Web (https://cdaweb.gsfc.nasa.gov/cdaweb/istp_public/), which is administrated by GSFC/NASA. We also thank the Helmholtz Centre Potsdam - GFZ German Research Centre for Geosciences and World Data Center for Geomagnetism, Kyoto for accessing the data of the $K_p$, $AU$, and $AL$ indices from https://www.gfz-potsdam.de/en/kp-index and http://wdc.kugi.kyoto-u.ac.jp/index.html.

**Figure 1**. Plots of solar wind parameters and geomagnetic activity indices during 5 h from 17:00 UT to 22:00 UT are shown. From top to bottom: the $AL$ and $AU$ indices, the IMF-$B_y$ and -$B_z$ components, the IMF clock angle (arctan(IMF-$B_y$/IMF-$B_z$)), and the Akasofu-Pelleaut parameter ($\varepsilon_{AP}$), obtained with $V_{SW}B_t^2\sin^4(\theta_{CLOCK}/2)(4\pi L_0^2/\mu_0)$, where $V_{SW}$ is the solar wind velocity, $B_t$ = sqrt(IMF-$B_x^2$ + IMF-$B_y^2$ + IMF-$B_z^2$), and $L_0$ = 7.0 $R_E$. The $K_p$ index is indicated at the bottom of the figure. The auroral spiral interval from 19:59 UT to 21:23 UT is indicated with horizontal thick bars and also bracketed with black broken vertical lines.

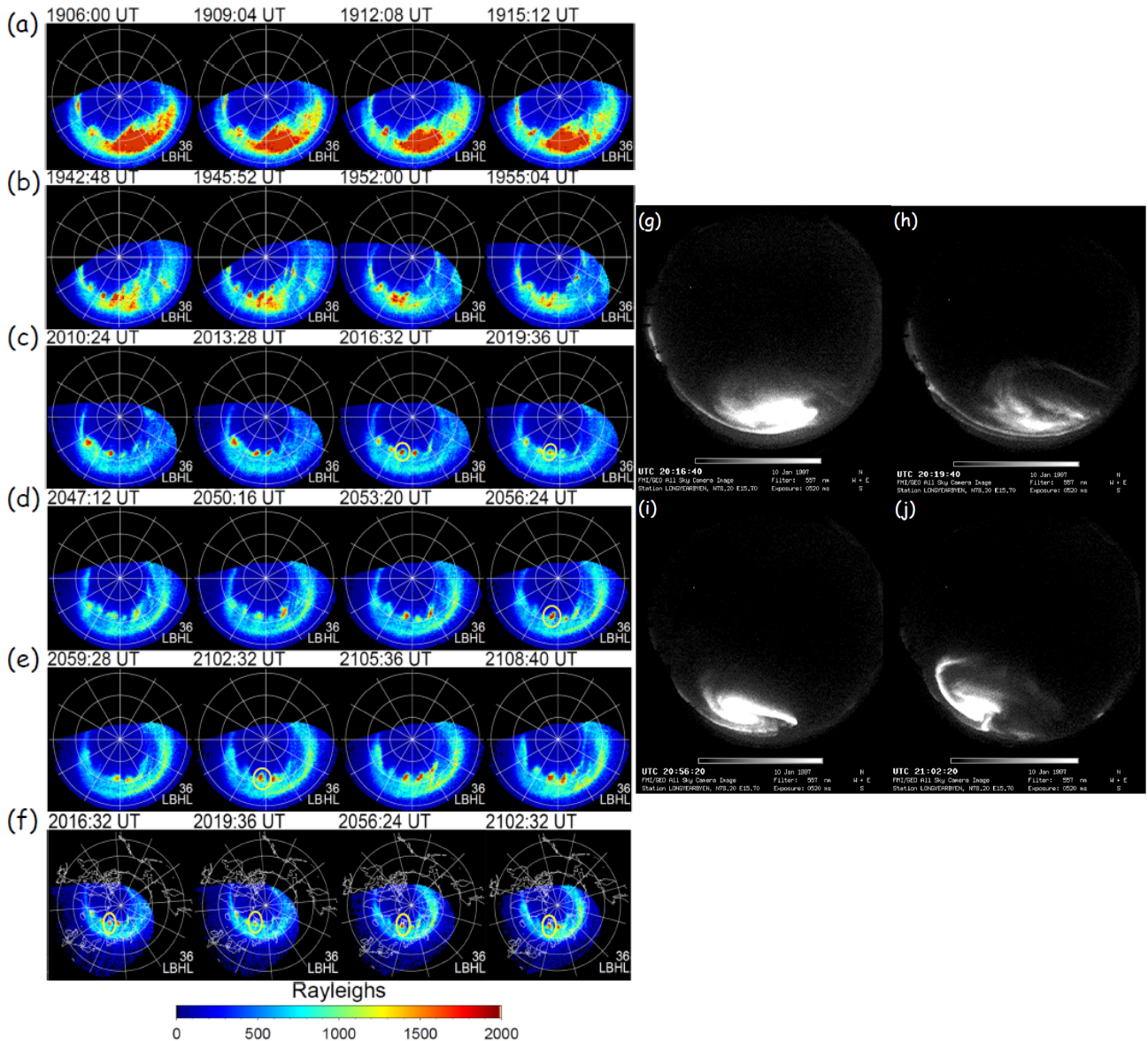

**Figure 2.** Consecutive Lyman-Birge-Hopfield long (LBHL) emission snapshots taken from the Polar ultraviolet imager (UVI) in altitude adjusted corrected geomagnetic (AACGM) coordinates (Baker and Wing, 1989) prior to (panels a and b) and during (panels c, d, and e) the auroral spiral are shown. The integration time of each image is 36.8 s. In panel f, four snapshots of LBHL-36s images for the nearest auroral spiral times, identified with the all-sky camera (ASC) images from Longyearbyen, are shown in geographic coordinates. Each panel is oriented such that the bottom, right, top, and left correspond to midnight (24 h magnetic local time; MLT), dawn (6 h MLT), noon (12 h MLT), and dusk (18 h MLT), respectively. The white circles are drawn every 10° from 60° to 80° magnetic latitude (MLat) for panels a – e and from 50° to 80° for panel f. The white lines are drawn every 2 h MLT. The color code is assigned according to the auroral brightness in units of Rayleigh. Panels g, h, i, and j show the images of the auroral spiral taken from the ASC installed at Longyearbyen. The approximate field of view of the ASC is marked with yellow circles in the UVI plots in panels c, d, e, and f.

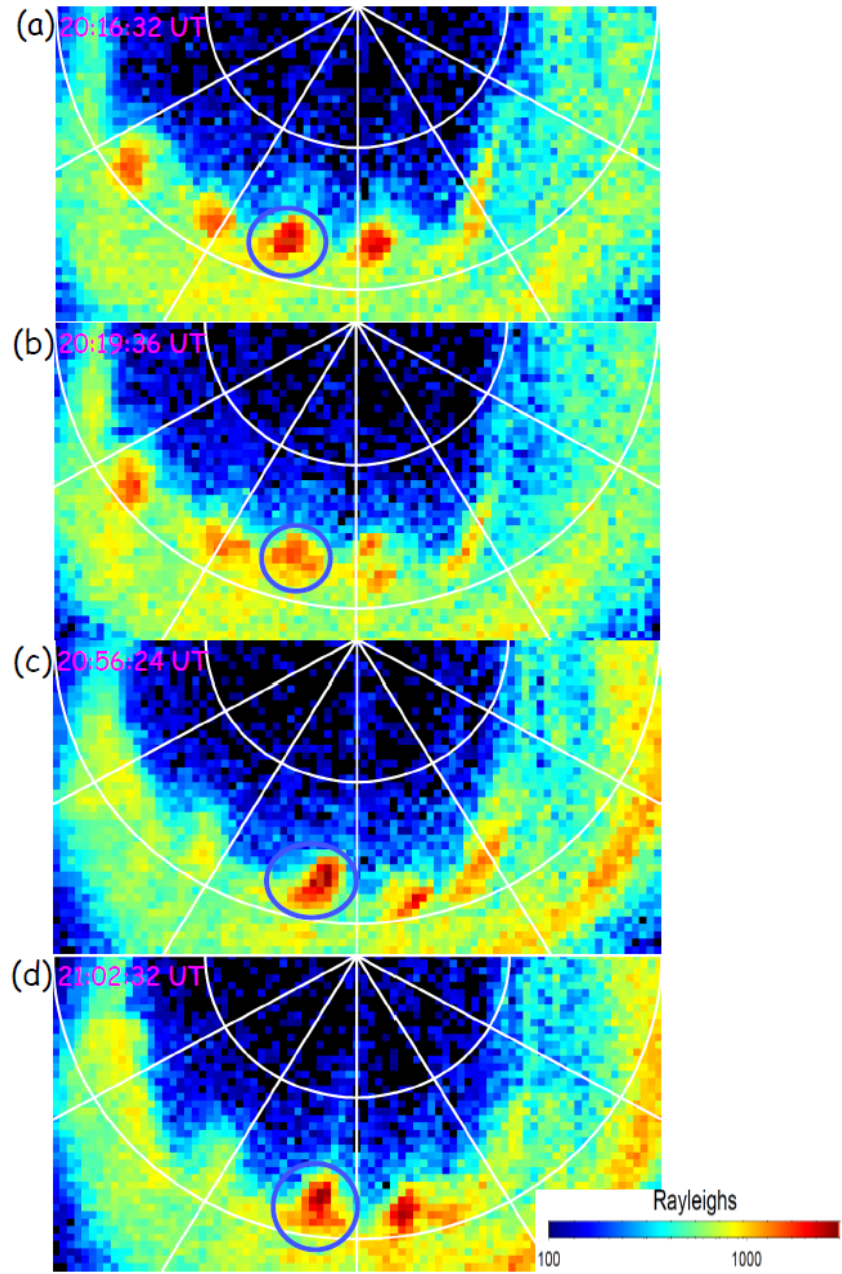

**Figure 3.** Zoomed-in snapshots of LBHL-36s images at the four auroral spiral times in the nightside from 18 h to 6 h MLT are shown. The time sequence of these images is the same as that of Figure 2f. The color code is assigned according to the logarithm of auroral brightness in units of Rayleigh. The white semicircles are drawn every 10° from 70° to 80° MLat, and the white lines are drawn every 2 h MLT. The blue circle indicates the auroral spiral which was detected by the ASC at Longyearbyen (LYR).

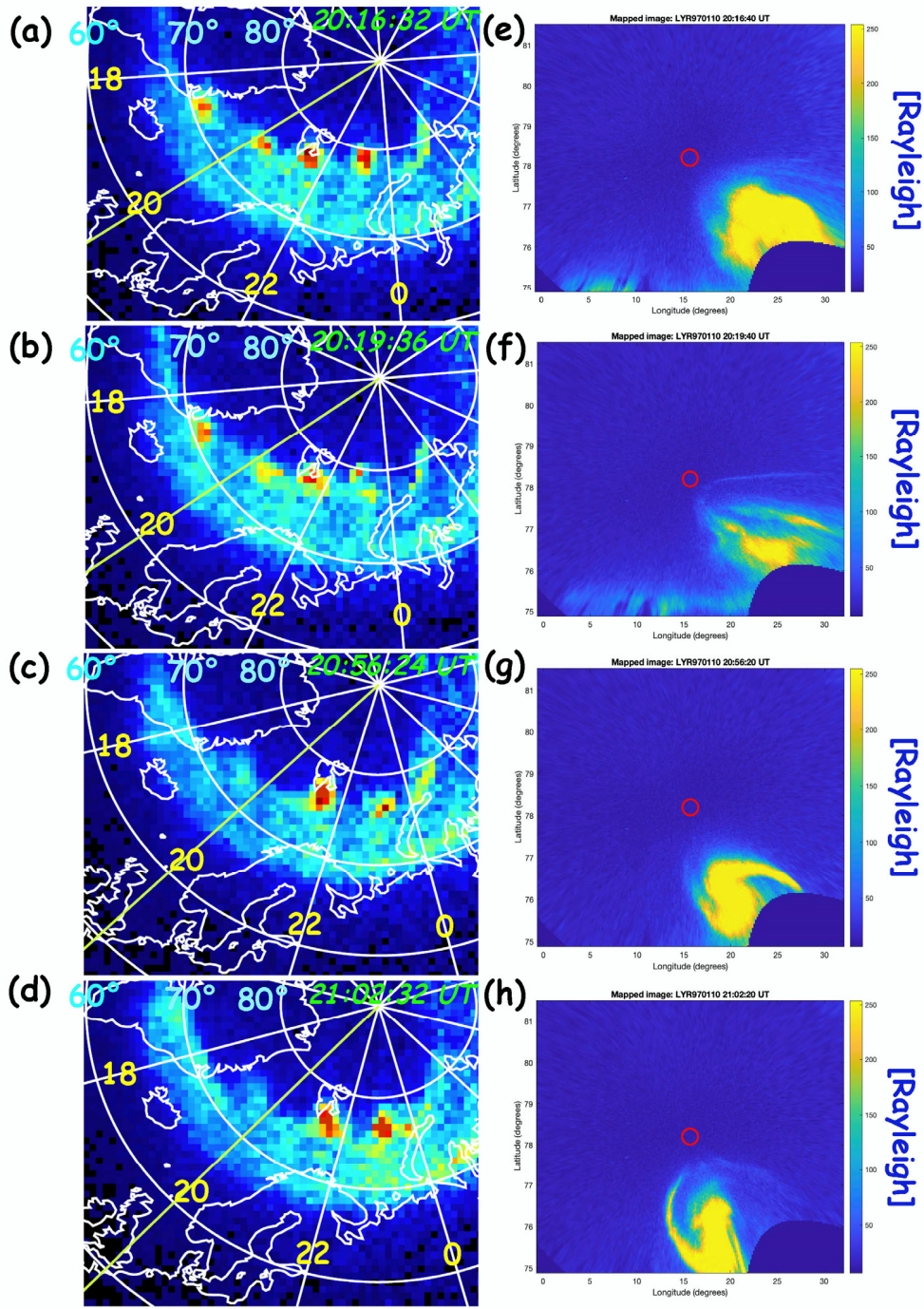

**Figure 4.** Zoomed-in snapshots of Figure 2f for the regions near Svalbard (Longyearbyen) are shown in panels a – d. The all-sky camera (ASC) data plots in geographic coordinates for four times are shown. Panels e – h show the plots for 20:16:40 UT, 20:19:40 UT, 20:56:20 UT, and 21:02:20 UT. The red circle in each panel indicates the location of the Longyearbyen station. In these plots, we assumed that the auroral emission height was 110 km. Vertical and horizontal axes

give the geographical latitude and longitude, respectively, in units of degree (°). The color code is assigned according to the auroral luminosity in units of Rayleigh.

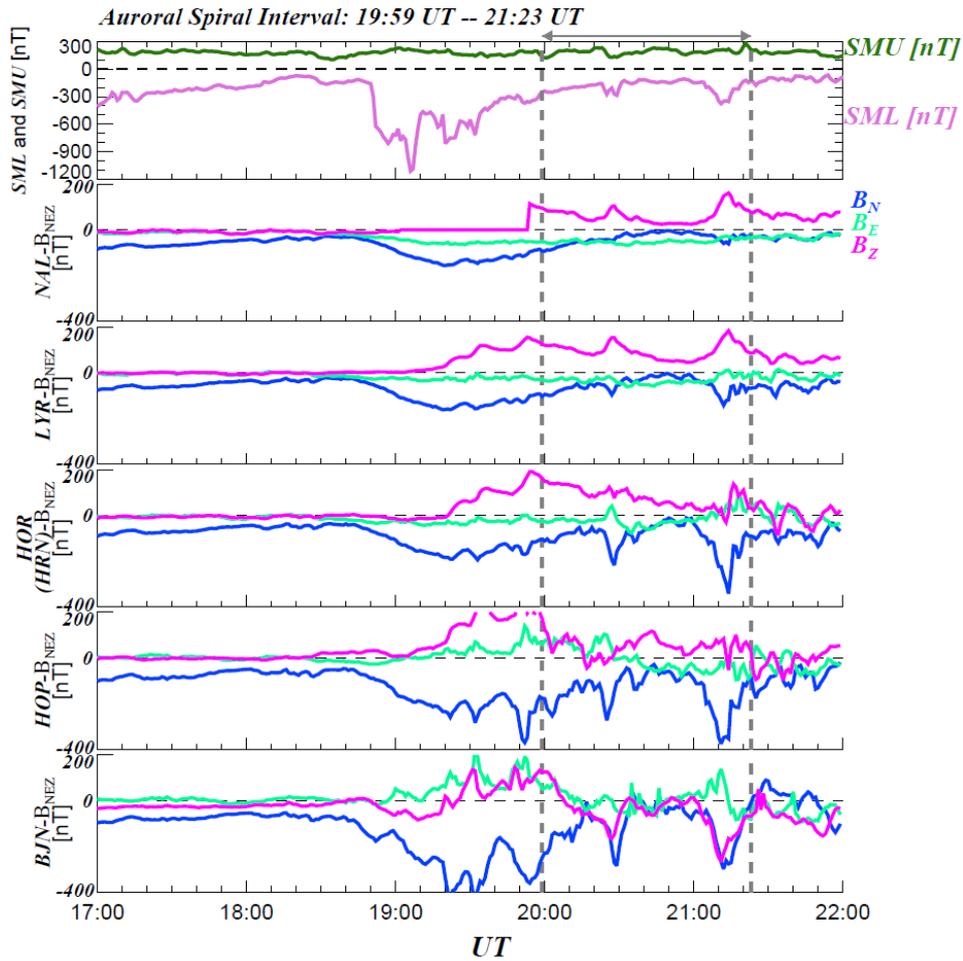

**Figure 5.** Plots of geomagnetic activity indices (*SML* and *SMU* indices) and geomagnetic field data obtained from the five magnetometers installed at and around the Longyearbyen station (LYR) on Svalbard island (the IMAGE geomagnetic observatory chain, also see Tanskanen, 2009) during 5 h from 17:00 UT to 22:00 UT are shown. From top to bottom panels display the *SML* and *SMU* indices, and the three components of the magnetic field ($B_N$, $B_E$, and $B_Z$) at the five magnetometer stations are shown. The N, E, and Z directions correspond to geomagnetic north-south, east-west, and up-down, respectively. These geomagnetic observatories have been sorted in order of decreasing latitude. The auroral spiral interval from 19:59 UT to 21:23 UT is indicated with a horizontal gray arrow and also bracketed with gray broken vertical lines.

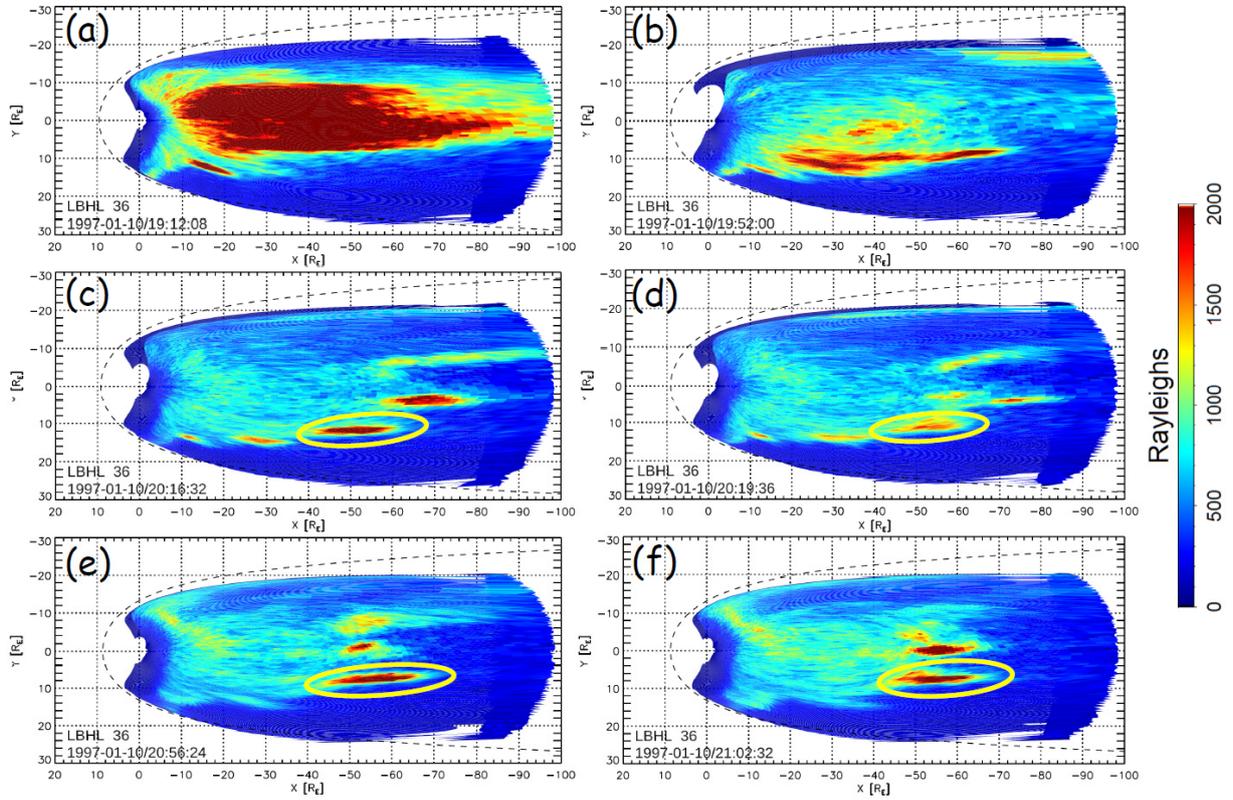

**Figure 6.** Magnetic equatorial projection maps of Polar UVI auroral pixel data prior to (panels a and b) and during (panels c, d, e, and f) the auroral spiral are shown. Each pixel was traced to the GSM X-Y plane of the magnetotail, based on the Tsyganenko 96 geomagnetic field model (Tsyganenko and Stern, 1996). Yellow ovals indicate the auroral spirals detected by the Polar UVI and the ASC at Longyearbyen. The color code shows the auroral brightness in units of Rayleigh. The dashed curve indicates the model magnetopause location derived from Shue et al. (1998).

**Table 1.** Rough scales (diameters) of the spiral core part (central part of intense auroral brightness) and the core part with the spiral arms are summarized. These scales are estimated from the ASC's images in geographical coordinates using 1° in geographical latitude = 111 km and the distance (km) corresponding to the geographical longitude where the auroral spiral was observed for each time. The calculations were performed with Latitude/Longitude Distance Calculator, developed by National Hurricane Center and Central Pacific Hurricane Center (https://www.nhc.noaa.gov/gccalc.shtml), by giving the geographical latitudes and longitudes at the two different edge points of the spiral.

| Time [UT] | Average scale of core part [km] | Average scale of core part + spiral arms [km] |
|-----------|--------------------------------|-----------------------------------------------|
| 20:16:40  | 250                            | 350                                           |
| 20:19:40  | 130                            | 300                                           |
| 20:56:20  | 180                            | 260                                           |
| 21:02:20  | 150                            | 320                                           |

## Auroral Morphological Changes to the Formation of Auroral Spiral during the Late Substorm Recovery Phase: Polar UVI and Ground All-Sky Camera Observations

Motoharu Nowada[1†], Yukinaga Miyashita[2,3], Noora Partamies[4,5],
Alexander William Degeling[1], and Quan-Qi Shi[1]

[1] Shandong Key Laboratory of Optical Astronomy and Solar-Terrestrial Environment, School of Space Science and Physics, Institute of Space Sciences, Shandong University, Weihai, Shandong, People's Republic of China.

[2] Korea Astronomy and Space Science Institute, Daejeon, South Korea.

[3] Korea University of Science and Technology, Daejeon, South Korea.

[4] Department of Arctic Geophysics, The University Centre in Svalbard, Norway.

[5] Birkeland Centre for Space Science, Norway.

## Contents of this file

Figures S1 and S2
Movie S1

## Introduction

Time-series of Polar ultraviolet imager (UVI) data of Lyman-Birge-Hopfield long (LBHL) emission with an integration time of 36.8 s is shown to examine the global auroral morphological changes before and during the formation of the aurora spiral from 19:59 UT to 21:23 UT on January 10, 1997. The UVI data are shown in two different coordinates: altitude adjusted corrected geomagnetic (AACGM) (Figure S1) and geographic coordinates (Figure S2) from 18:41:28 UT to 21:57:44 UT. All UVI data were obtained from the Northern Hemisphere. Here, we also show a movie of consecutive images obtained from ASC installed at the Longyearbyen station for 2 h from 20:00 UT to 22:00 UT, when the auroral spiral interval is covered (Movie S1).

Figure S1

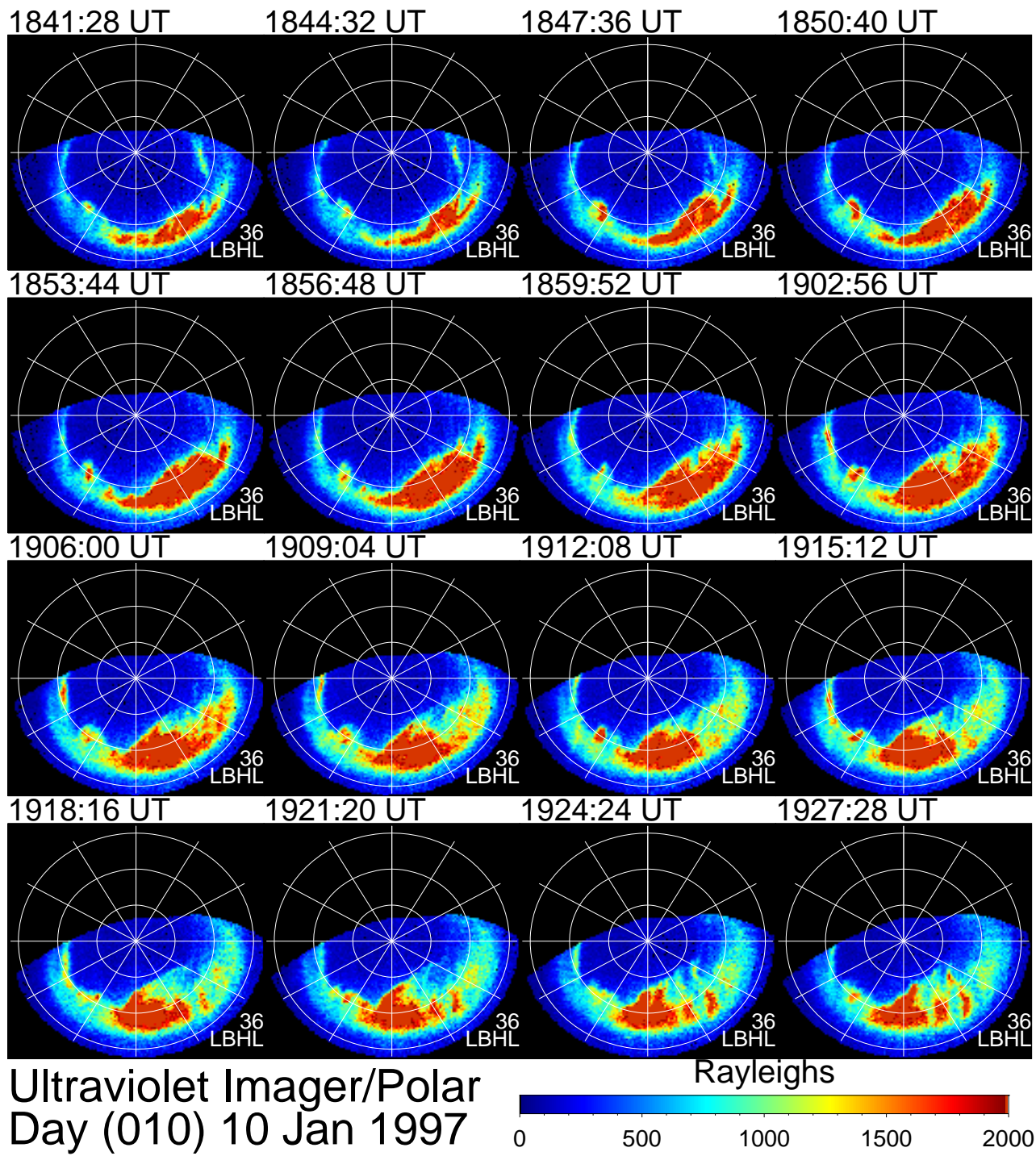

Ultraviolet Imager/Polar
Day (010) 10 Jan 1997

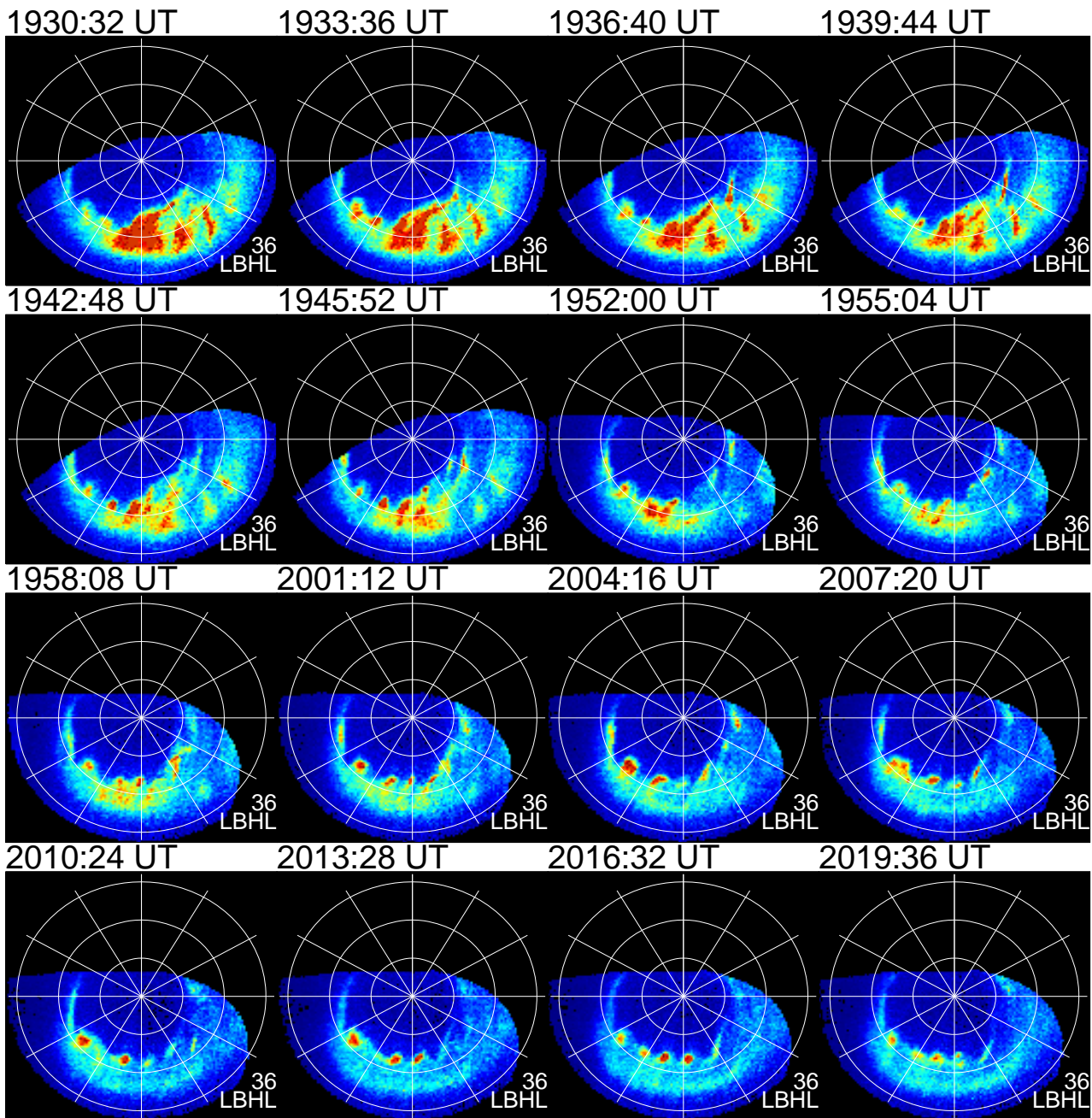

1930:32 UT 1933:36 UT 1936:40 UT 1939:44 UT
1942:48 UT 1945:52 UT 1952:00 UT 1955:04 UT
1958:08 UT 2001:12 UT 2004:16 UT 2007:20 UT
2010:24 UT 2013:28 UT 2016:32 UT 2019:36 UT

Ultraviolet Imager/Polar
Day (010) 10 Jan 1997

Rayleighs

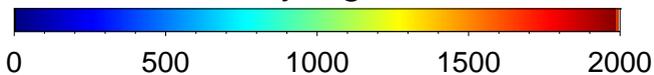

0    500    1000    1500    2000

Ultraviolet Imager/Polar
Day (010) 10 Jan 1997

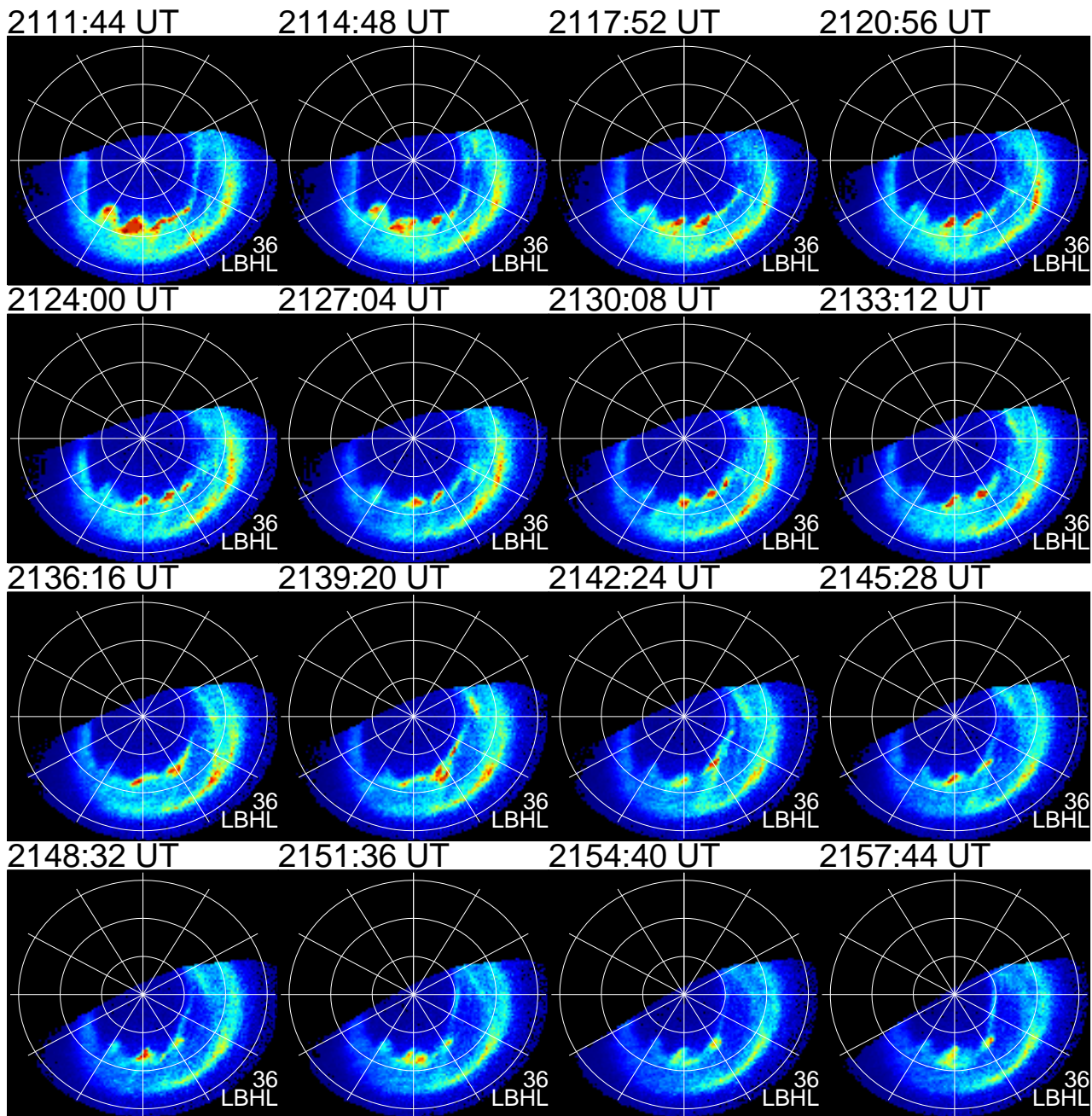

2111:44 UT   2114:48 UT   2117:52 UT   2120:56 UT

2124:00 UT   2127:04 UT   2130:08 UT   2133:12 UT

2136:16 UT   2139:20 UT   2142:24 UT   2145:28 UT

2148:32 UT   2151:36 UT   2154:40 UT   2157:44 UT

Ultraviolet Imager/Polar
Day (010) 10 Jan 1997

Rayleighs

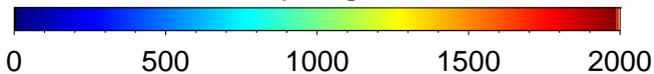

0    500    1000    1500    2000

Figure S2

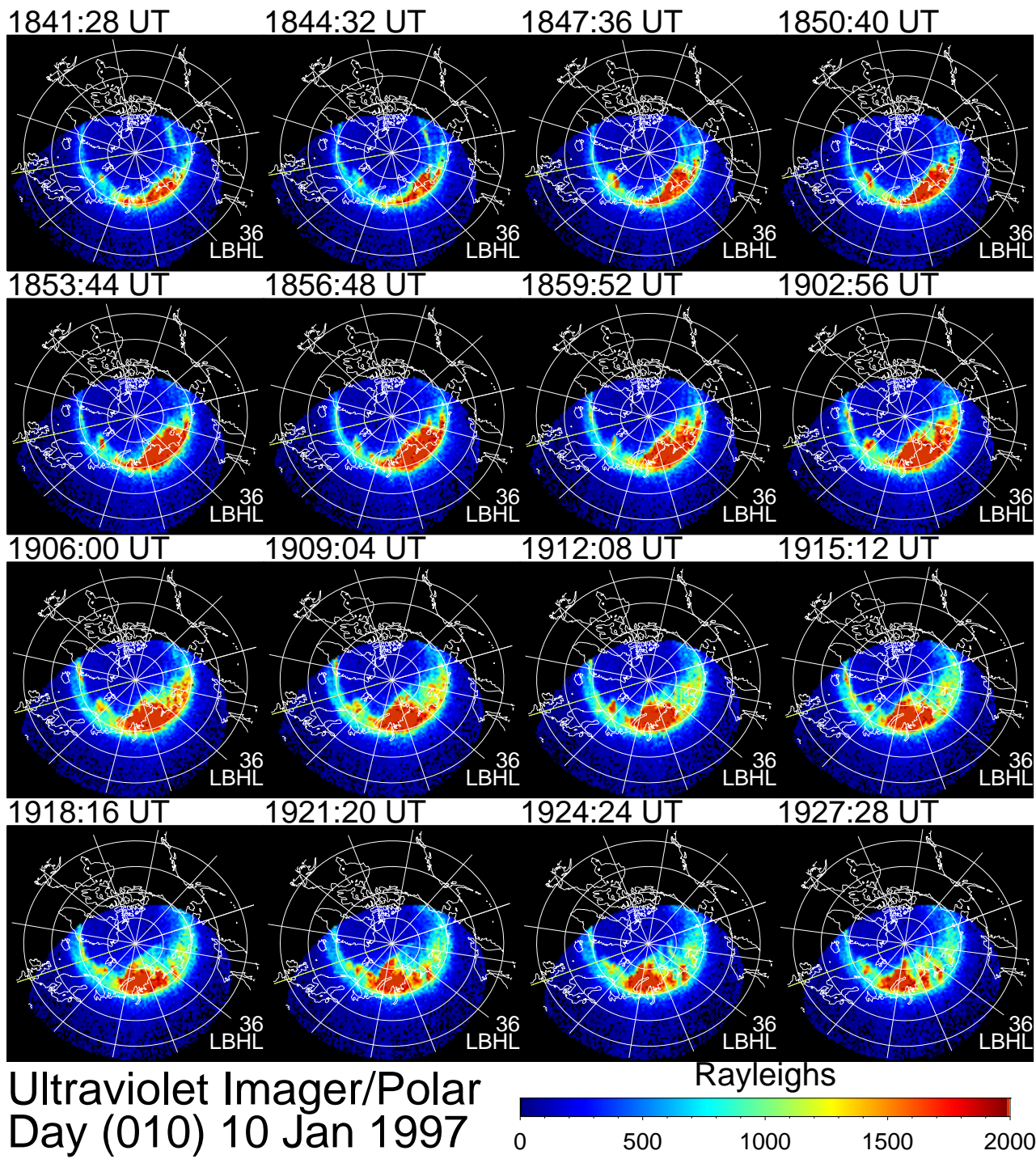

Ultraviolet Imager/Polar
Day (010) 10 Jan 1997

Rayleighs

0    500    1000    1500    2000

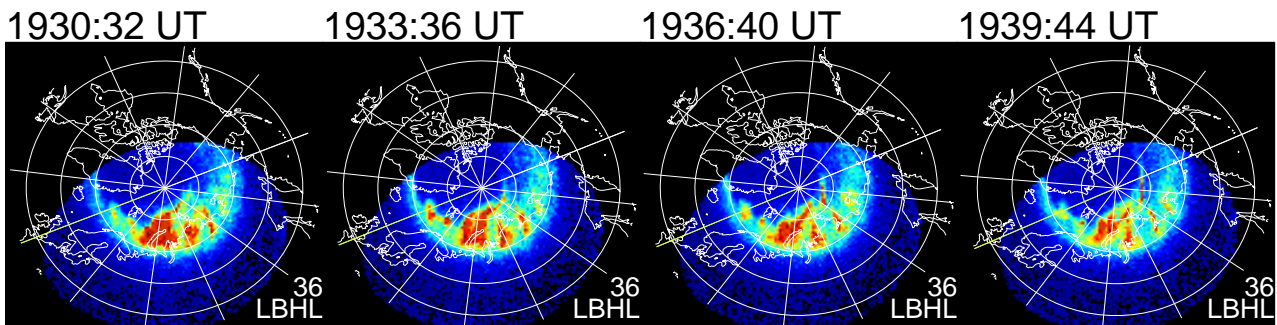

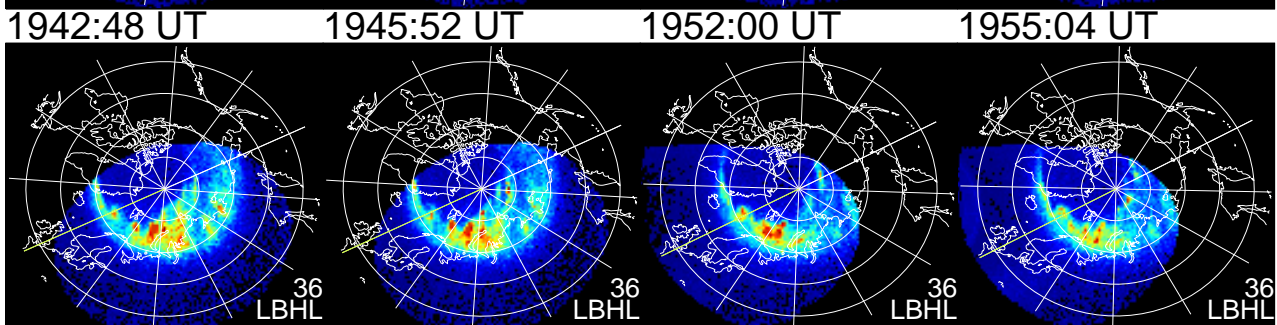

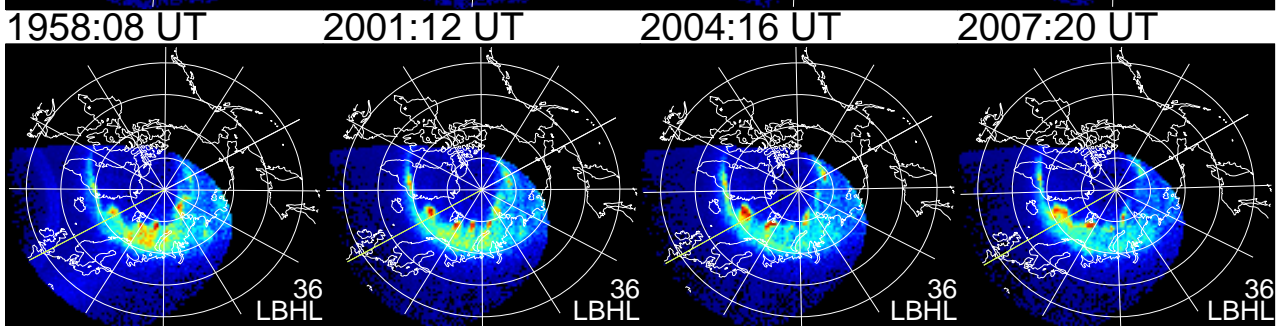

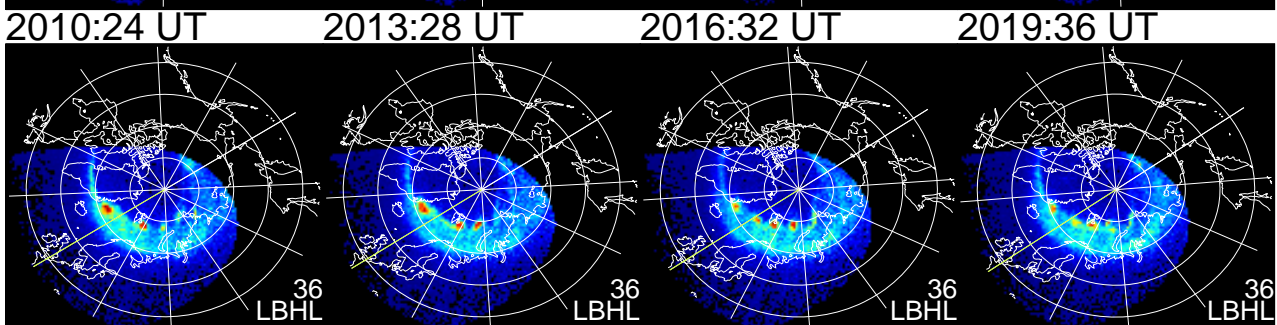

Ultraviolet Imager/Polar
Day (010) 10 Jan 1997

Rayleighs

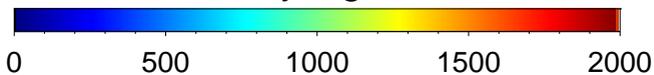

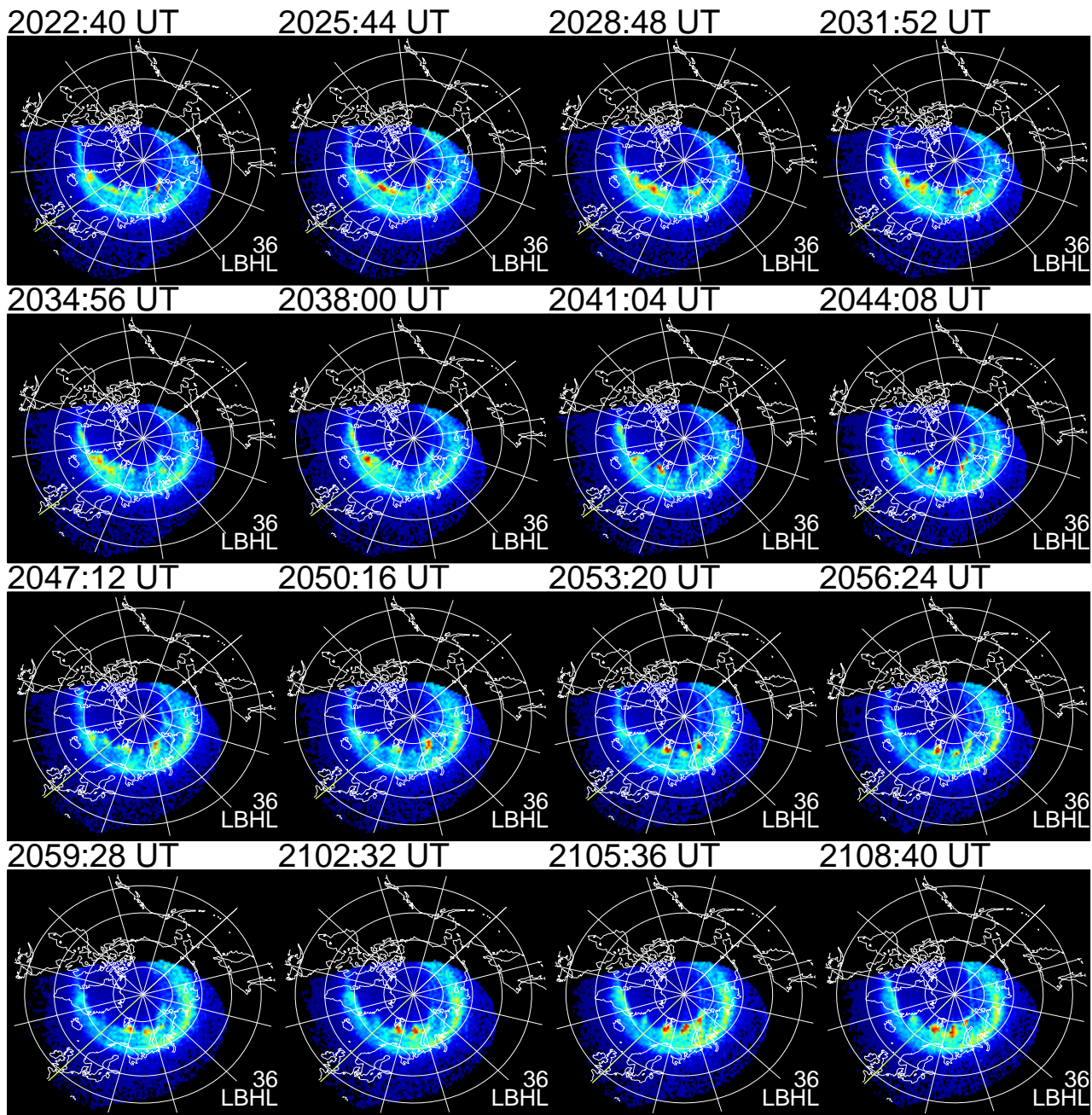

# Ultraviolet Imager/Polar
## Day (010) 10 Jan 1997

Rayleighs

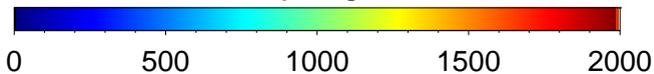

0   500   1000   1500   2000

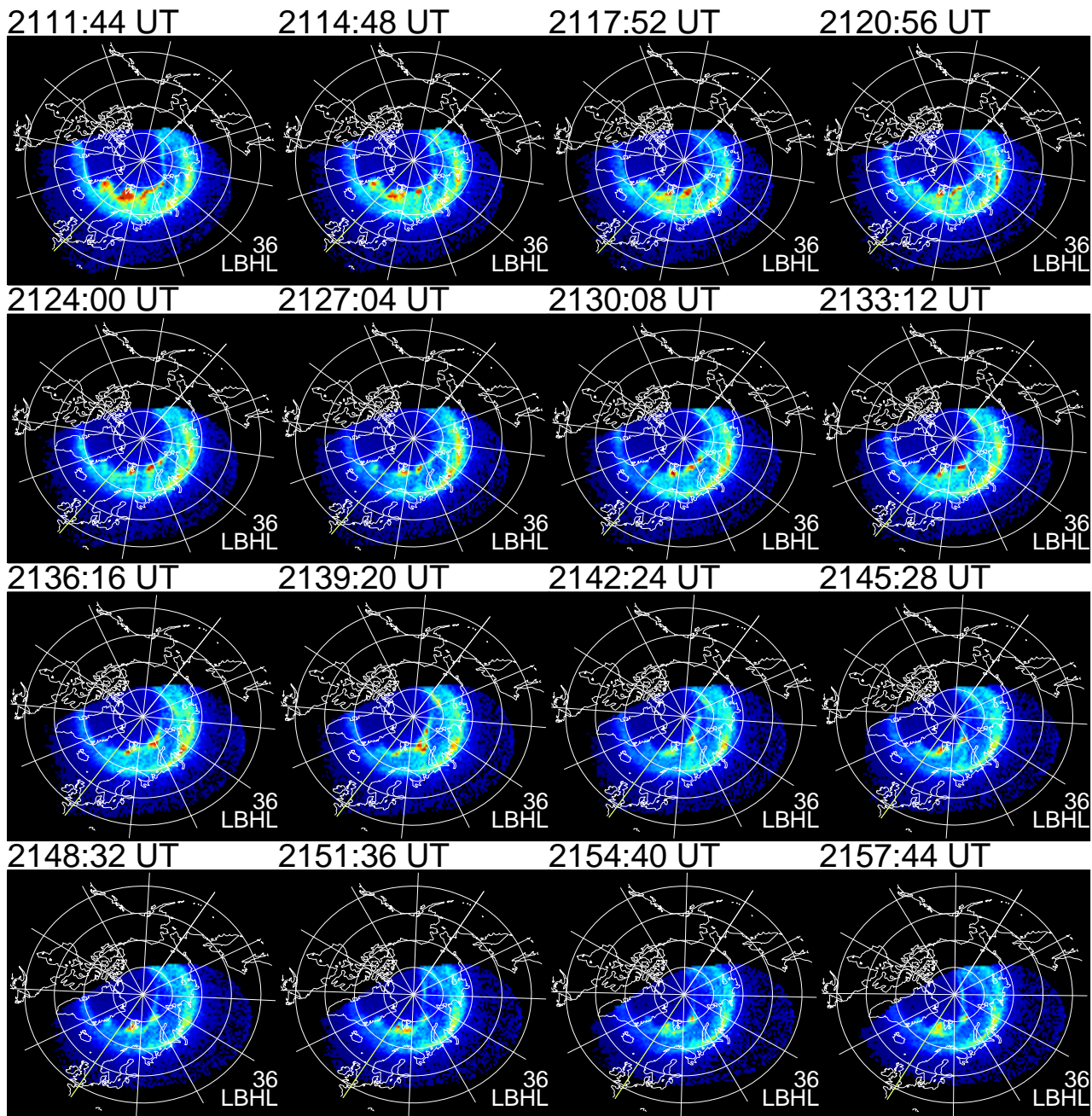

2111:44 UT    2114:48 UT    2117:52 UT    2120:56 UT

2124:00 UT    2127:04 UT    2130:08 UT    2133:12 UT

2136:16 UT    2139:20 UT    2142:24 UT    2145:28 UT

2148:32 UT    2151:36 UT    2154:40 UT    2157:44 UT

Ultraviolet Imager/Polar
Day (010) 10 Jan 1997

Rayleighs

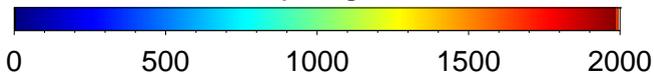

0    500    1000    1500    2000

**Figure S1.** Time-series snapshots of Polar ultraviolet imager (UVI) of Lyman-Birge-Hopfield long (LBHL) emission with an integration time of 36.8 s in altitude adjusted corrected geomagnetic (AACGM) coordinates are shown for 3 h 16 min from 18:41:28 UT to 21:57:44 UT. The white circles are drawn every 10° from 60° to 80° magnetic latitude (MLat). Each panel is oriented such that the top, right, bottom and left sides correspond to noon (12h MLT (magnetic local time)), dawn (6h MLT), midnight (24h MLT), and dusk (18h MLT), respectively. The white lines are drawn every 2 h MLT. The color code is assigned according to the luminosity in units of Rayleigh.

**Figure S2.** Time-series of Polar UVI snapshots in geographic coordinates with coast lines for the same time interval as Figure S1 are shown in the same format as Figure S1.

**Movie S1.** Movie of consecutive images of the aurora spiral for 2 h from 20:00 UT to 22:00 UT on January 10, 1997, obtained from the all-sky camera (ASC) installed at Longyearbyen on Svalbard island is shown. The ASC covers a circular area with a diameter of about 600 km at an altitude of 110 km, with a field of view of 140°. Because the number of pixels corresponding to a 140° field of view is 440×440, the average spatial resolution of the image is about 1.4 km/pixel. The temporal resolution of the ASC's images is 20 s. The top, right, bottom, and left correspond to north, east, south, and west, respectively.